\documentclass[a4paper,12pt]{article}
\usepackage[T2A]{fontenc}
\usepackage[utf8]{inputenc}   % older versions of ucs package
\usepackage[russian,english]{babel}

\usepackage{verbatim} 	% to be able to use \begin{comment} ... \end{comment}
\usepackage{amsmath, amsthm}
\usepackage{graphicx}
\usepackage{array} 	% needed for \setlength{\extrarowheight}{0.3cm} 
\usepackage{rotating}	% This allows you to rotate text, tables and figures in every manner.

\usepackage{hyperref}	% for hreferences. Use like this in document:
\begin{document}
\pagestyle{plain}
\pagenumbering{arabic}

\selectlanguage{english}

\title{The effects of multiply quantum wells (MQW) on optical and electrical characteristics of AlGaAs 
lasers with separate confinement heterostructures (SCH).}

\date{}

\author{Z. Koziol\footnote{Corresponding author email: zbigniew@ostu.ru} ~and S. I. Matyukhin\\
Orel State Technical University,\\
29 Naugorskoye Shosse, Orel, 302020, Russia.}

\maketitle

\abstract{
Optical and electrical characteristics of AlGaAs lasers with separate confinement heterostructures
are modeled by using Synopsys's Sentaurus TCAD, and open source software. 
The results for cases of 2-QW (2 Quantum Wells) 
and 3-QW structures are compared with these for 1-QW. A significant improvement of useful laser
parameters is obtained by increasing the number of Quantum Wells and optimizing the width of waveguides.
In particular, the maximum optical efficiency is shown to
reach 88 \% for a 3-QW structure with optimal width of waveguides. 
The width of optical intensity profile of MQW lasers increases, 
leading to lowering maximal light power density passing through laser facets, decreasing the 
risk of catastrophic damage of mirrors.
}

\section{Introduction}

Alferov \cite{Alferov}, et al., proposed creating semiconductor-based lasers
comprising the use of a geometrically-narrow active recombination region 
where photon generation occurs, 
with waveguides around improving the gain to loss ratio
(separate confinement heterostructures; SCH). That idea dominated largely 
the field of optoelectronics development
in the past years. Due to the relative simplicity and perfection of technology,
solid solutions of $Al_x Ga_{1-x}As$ are commonly used as wide-gap semiconductors in SCH lasers.

Finding however the best, most efficient structures of SCH lasers is not a task that could be solved
easy by experimenting only. Computational modeling is being used recently,
as the cost effective approach towards designing new devices as well predicting their characteristics.
Moreover, computational modeling offers also new approaches to study physical properties of a device: 
often, limits of technology do not allow to carry out experiments with accuracy high enough that would
allow to verify certain hypothesis.

In our earlier works we first were able to find agreement between our calculations of quantum 
well energy states and the lasing wavelength observed experimentally \cite{Koziol}. 
Next \cite{Koziol_doping}, several changes in structure and doping of SCH AlGaAs lasers 
have been shown to considerably improve their electrical and optical parameters. 
We compared computed properties with these of lasers produced by Polyus research 
institute in Moscow \cite{Andrejev}, \cite{Andrejev_2}.
In particular, by changing the width of active region (Quantum Well), waveguide width, doping 
concentration in all laser layers, as well by changing the waveguide profile by introducing a gradual
change of Al concentration, and also by introducing variable doping profile of carriers across 
waveguide, we were able to decrease significantly the lasing threshold current, 
increase the slope of optical power versus current, and increase optical efficiency up to about 
74 \% \cite{Waveguide}. 
We have shown \cite{NGC} that the lasing action may not occur at certain widths
or depths of Quantum Well (QW), and the threshold current as a function of these parameters may have 
discontinuities that occur when the most upper quantum well energy values are very close to either 
conduction band or valence band energy offsets. These effects are more pronounced at low temperatures, 
and may be observed also, at certain conditions, in temperature dependence of lasing threshold current 
as well.

A simple analytical, phenomenological model was shown to describe optical efficiency, 
$\eta=L/P$ ($L$ is optical power intensity and $P$ is the supplied electrical power), 
with a high accuracy, by using two parameters only \cite{differential}.

In Multi-Quantum Well (MQW) structures better characteristics are achievable than
in case when one QW is used. Hence, a question is to what an extend 
yet could we improve properties of laser studied, by using optimized results obtained already 
and by introducing MQW. The purpose of this work is to compare results for 1, 2, and 3 QW structures
and find out optimal widths of waveguides in each case.

One of the fundamental laser characteristics are their threshold current, $I_{th}$, above which 
lasing action starts, the slope of optical power versus current, $S=dL/dI$, 
which is approximately constant for currents just above $I_{th}$, and lasing offset voltage $U_0$. 
We choose these parameters as characteristics to be compared in modeling.

\section{Lasers structure and calibration of modeling.}

\begin{table}[t]
\caption{Reference structure of AlGaAs 1-QW SCH laser layers used in computer modeling. \emph{d} is the width of layers.}
\label{table_1}
%\begin{center}
%\begin{tabular}{|c|c|c|c|c|c|}
  \begin{tabular}[htbp]{@{}lllll@{}}
\hline
No  &       Layer&    Composition&    Doping [$cm^{-3}$]&    d [$\mu m$]\\
\hline
1        & n-substrate     &   n-GaAs (100)    &  $2\cdot 10^{18}$  &    350              \\
%\hline
2        & n-buffer        &   n-GaAs           &  $1\cdot 10^{18}$ &    0.4              \\
%\hline
3        & n-emitter       &   $Al_{0.5}Ga_{0.5}As$ & $1\cdot 10^{18}$ &    1.6              \\
%\hline
4        & waveguide       &   $Al_{0.33}Ga_{0.67}As$ & $n \approx 10^{15}$ &  0.2                \\
%\hline
5        & QW 					& $Al_{0.08}Ga_{0.92}As$ & $n \approx 10^{15}$ &  0.012                \\
%\hline
6        & waveguide       &   $Al_{0.33}Ga_{0.67}As$ & $n \approx 10^{15}$ &  0.2                \\
%\hline
7        & p-emitter       &   $Al_{0.5}Ga_{0.5}As$ &  $1\cdot 10^{18}$ &  1.6        \\
%\hline
8        & contact		   &   p-GaAs           & $4\cdot 10^{19}$ &    0.5      \\
\hline
\end{tabular}
%\end{center}
\end{table}

\begin{table}[t]
\caption{Summary of experimental conditions and laser parameters, 
for the reference structure described in Table \ref{table_1}.}
\label{table_2}
%\begin{center}
%\begin{tabular}{|c|c|c|c|c|c|}
\begin{tabular}[htbp]{@{}lllll@{}}
\hline
Temperature [K] &   300       \\
%\hline
Lasing wavelength [nm]&   808 \\
%\hline
Offset voltage $U_0$ [V] &   1.56-1.60\\
%\hline
Differential resistance, $r=dU/dI$  [m$\Omega$] &   50-80   \\
%\hline
Threshold current $I_{th}^0$ [mA] & 200-300 \\
%\hline
Slope of optical power, $S_0=dL/dI$ [W/A]&  1.15-1.25 \\
%\hline
Left mirror reflection coefficient $R_l$ &  0.05 \\
%\hline
Right mirror reflection coefficient $R_r$ &  0.95 \\
\hline
\end{tabular}
%\end{center}
\end{table}

\begin{table}[h]
\caption{Two Quantum Wells: Structure of AlGaAs SCH laser layers. \emph{d} is the width of layers.}
\label{table_3}
%\begin{center}
%\begin{tabular}{|c|c|c|c|c|c|}
  \begin{tabular}[htbp]{@{}lllll@{}}
\hline
No  &       Layer&    Composition&    Doping [$cm^{-3}$]&    d [$\mu m$]\\
\hline
1        & n-emitter       &   $Al_{0.5}Ga_{0.5}As$ & $N~~1\cdot 10^{18}$ &    1.6              \\
%\hline
2        & waveguide  &   $Al_{0.33}Ga_{0.67}As$ & $N~~1\cdot 10^{15}$ &  $d_{w0}$                \\
%\hline
3        & QW & $Al_{0.08}Ga_{0.92}As$ & $N~~1\cdot 10^{15}$ &  0.012                \\
%\hline
4        & waveguide &   $Al_{0.33}Ga_{0.67}As$ & $N~~5\cdot 10^{14}$ &  $d_{w1}$                \\
%\hline
5        & QW & $Al_{0.08}Ga_{0.92}As$ & $P~~1\cdot 10^{15}$ &  0.012                \\
%\hline
6        & waveguide       &   $Al_{0.33}Ga_{0.67}As$ & $P~~1\cdot 10^{15}$ &  $d_{w0}$                \\
%\hline
7        & p-emitter       &   $Al_{0.5}Ga_{0.5}As$ &  $P~~1\cdot 10^{18}$ &  1.6        \\
\hline
\end{tabular}
%\end{center}
\end{table}

\begin{table}[h]
\caption{Three Quantum Wells: Structure of AlGaAs SCH laser layers. \emph{d} is the width of layers.}
\label{table_4}
%\begin{center}
%\begin{tabular}{|c|c|c|c|c|c|}
  \begin{tabular}[htbp]{@{}lllll@{}}
\hline
No  &       Layer&    Composition&    Doping [$cm^{-3}$]&    d [$\mu m$]\\
\hline
1        & n-emitter       &   $Al_{0.5}Ga_{0.5}As$ & $N~~1\cdot 10^{18}$ &    1.6              \\
%\hline
2        & waveguide  &   $Al_{0.33}Ga_{0.67}As$ & $N~~1\cdot 10^{15}$ &  $d_{w0}$                \\
%\hline
3        & QW & $Al_{0.08}Ga_{0.92}As$ & $N~~1\cdot 10^{15}$ &  0.012                \\
%\hline
4        & waveguide &   $Al_{0.33}Ga_{0.67}As$ & $N~~5\cdot 10^{14}$ &  $d_{w1}$                \\
%\hline
5        & QW & $Al_{0.08}Ga_{0.92}As$ & $N~~5\cdot 10^{14}$ &  0.012                \\
%\hline
6        & waveguide &   $Al_{0.33}Ga_{0.67}As$ & $P~~5\cdot 10^{14}$ &  $d_{w1}$                \\
%\hline
7        & QW & $Al_{0.08}Ga_{0.92}As$ & $P~~5\cdot 10^{14}$ &  0.012                \\
%\hline
8        & waveguide       &   $Al_{0.33}Ga_{0.67}As$ & $P~~1\cdot 10^{15}$ &  $d_{w0}$                \\
%\hline
9        & p-emitter       &   $Al_{0.5}Ga_{0.5}As$ &  $P~~1\cdot 10^{18}$ &  1.6        \\
\hline
\end{tabular}
%\end{center}
\end{table}

\begin{figure}[h]
\begin{center}
      \resizebox{150mm}{!}{\includegraphics{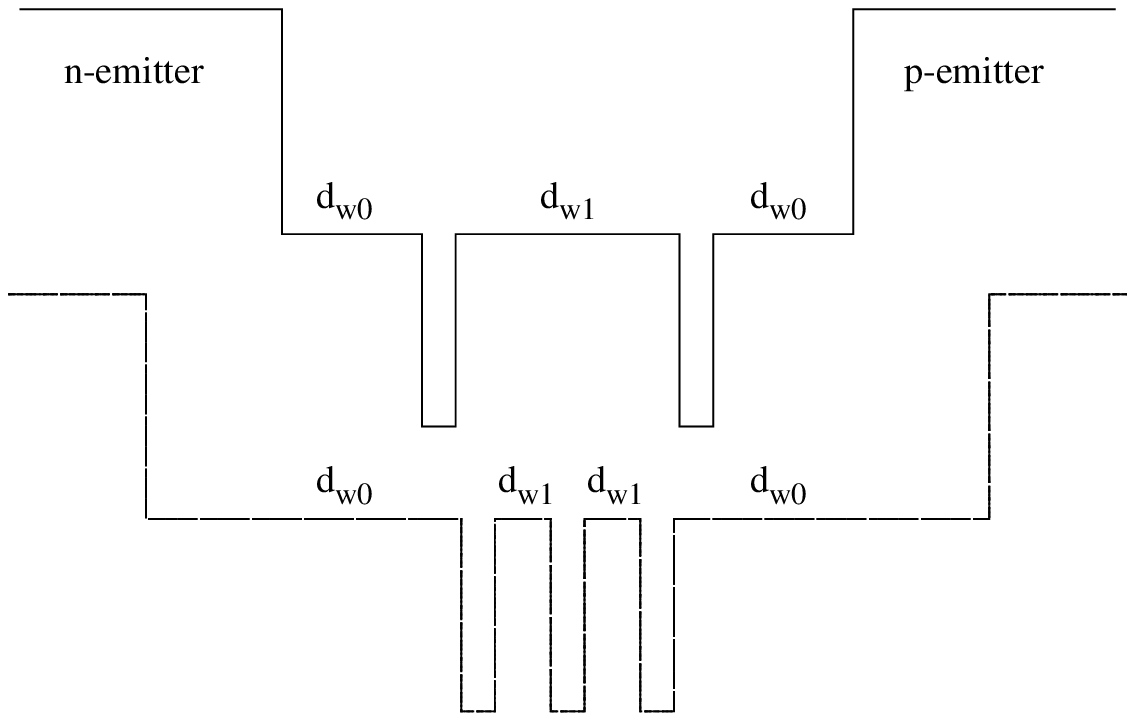}}
      \caption{Schematic structure of energy gap for MQW lasers modeled, with 2- and 3-quantum wells 
	(the upper and lower diagrams, respectively). $d_{w0}$ and $d_{w1}$ 
	are widths of waveguide regions. Active regions remain the same and their width is $12 nm$.
	Doping types and concentrations are shown in Tables \ref{table_3} and \ref{table_4}.
}
      \label{structure_00}
\end{center}
\end{figure}

The reference laser we model has a $1000 \mu m$ cavity length and $100 \mu m$
width, with doping and Al-content as described in Table \ref{table_1}.
The Table \ref{table_2} describes it's experimental parameters.
The structure of 2-QW and 3-QW lasers, as shown schematically in Figure \ref{structure_00},
is described in Tables \ref{table_3} and \ref{table_4}, respectively.

Synopsys's Sentaurus TCAD is used for modeling \cite{tcad}. This is an advanced, flexible commercial
computational environment used for modeling a broad range of technological and physical 
processes in microelectronics world. 
In case of lasers, some calculations in Sentaurus have purely phenomenological nature. 
The electrical and optical characteristics depend, primarily, on the following
computational parameters that are available for adjusting:

AreaFactor of electrodes, $A_e$,  AreaFactor in Physics section, $A_{ph}$, 
electrical contact resistance $R_x$, and reflection coefficient of laser mirrors,
$R_l$ and $R_r$. There are several parameters for adjustment that are related to microscopic
physical properties of materials or structures studied. However, often their values are either 
unknown exactly or finding them would require quantum-mechanical modeling based on first-principles.
This is however not the aim of our work.

In order to find agreement between the calculated results and these observed experimentally 
we adjust accordingly values of $A_e$ and $A_{ph}$. 

The results for $I_{th}$ and $S$ are all shown normalized by the values for the reference
laser described in Table \ref{table_1}, $I_{th}^0$ and $S_0$, respectively.

We neglect here the effect of contact resistance, 
by not including buffer and substrate layers and contacts into calculations
(compare with structure described in Table \ref{table_1}). 
We use instead InnerVoltage parameter available in Sentaurus and treat it as a physical quantity 
that is related to voltage applied. However, we examined in details results of calculations
of optical and electrical characteristics and compared them with these from measurements, finding 
a reasonably good agreement between them, for lasing offset voltage, threshold current, optical intensity, 
and differential resistivity \cite{differential}.

%%%%%%%%%%%%%%%%%%%%%%%%%%%%%%%%%%%%%%%%%%%%%%%%%%%%%%%%%%%%%%%%%%%%%%%%%%%%%%%%%%%%%%%%%%%
\section{Threshold current and $dL/dI$ for 2-QW and 3-QW structures.}

We choose to use threshold current $I_{th}$ and the slope of optical intensity versus current, $S=dL/dI$ 
as the most important technologically laser parameters, when comparing lasers with various structures. 
The advantage of using them is also in that that these could be easy extracted semi-automatically
from a large collection of data sets. Details of data analysis are described better in \cite{differential} 
and also on our laboratory web site\footnote{Web: www.ostu.ru/units/ltd/zbigniew/synopsys.php}.

Computation was performed as a function of the width $d_{w1}$, for a certain set of constant values of $d_{w0}$ 
(see Fig. \ref{structure_00}). 

\begin{figure}[h]
\begin{center}
      \resizebox{150mm}{!}{\includegraphics{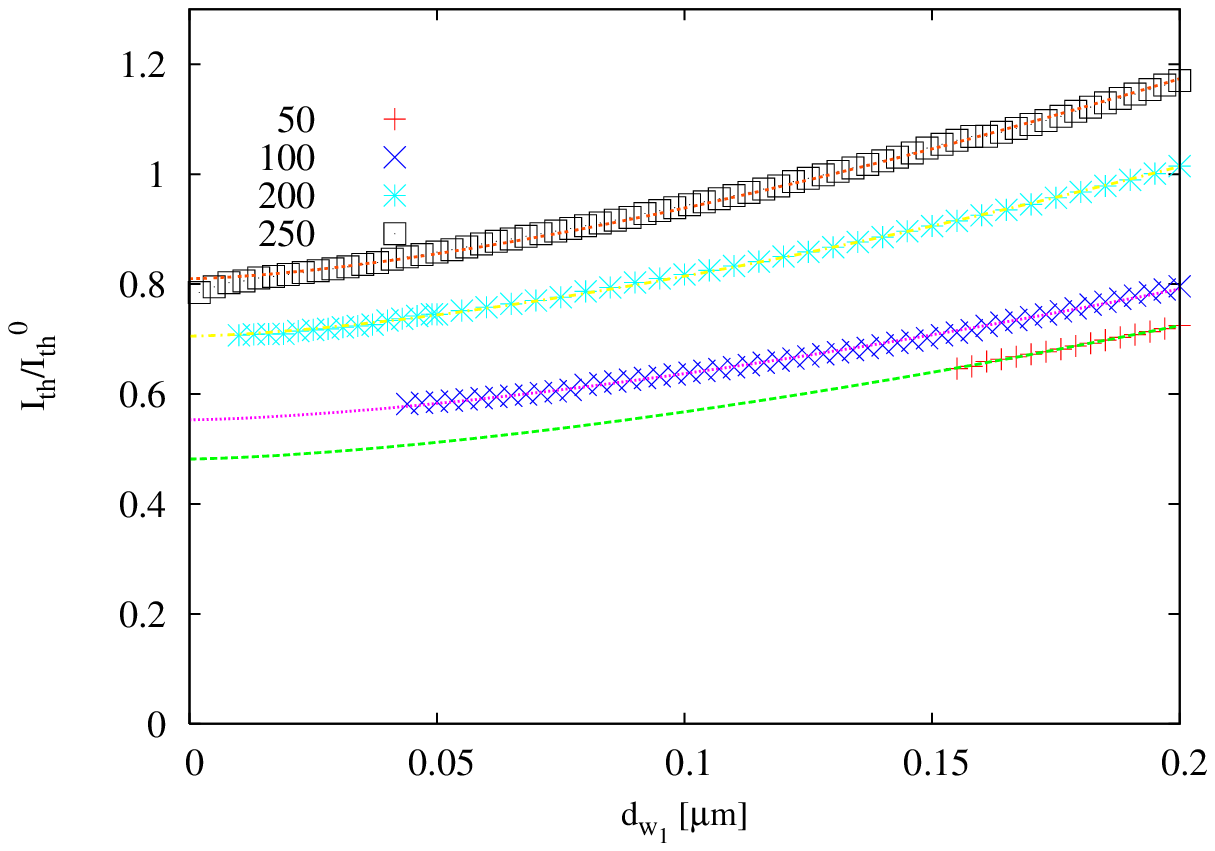}}
      \caption{Two Quantum Wells: threshold current, $I_{th}$, normalized by $I_{th}^0$, 
	as a function of the width of waveguide in the center, for a few width values 
	of other two waveguides (in $nm$), as indicated in the Figure.
}
      \label{2active_00}
\end{center}
\end{figure}

\begin{figure}[h]
\begin{center}
      \resizebox{150mm}{!}{\includegraphics{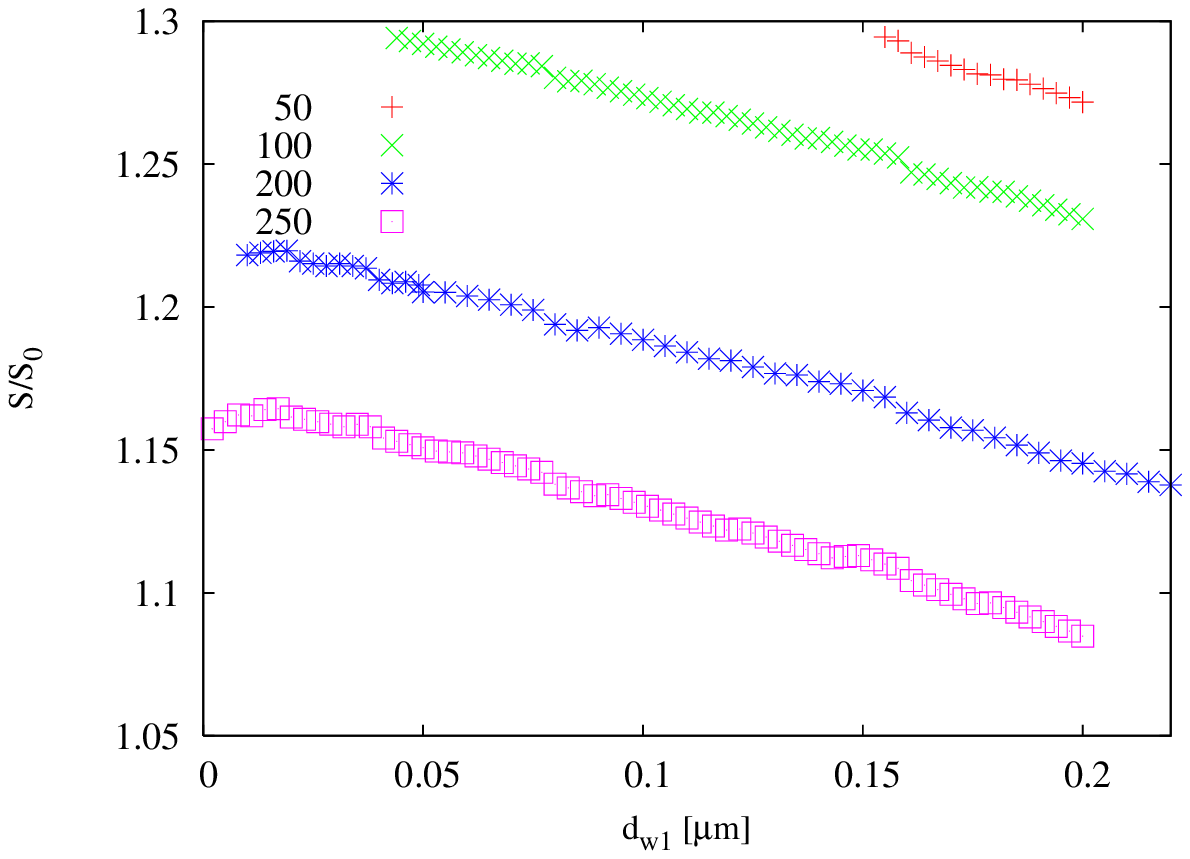}}
      \caption{Two Quantum Wells: Slope of optical power, $S=dL/dI$, normalized by $S_0$, 
	corresponding to the data in Figure \ref{2active_00}.
}
      \label{2active_01}
\end{center}
\end{figure}

\begin{figure}[h]
\begin{center}
      \resizebox{150mm}{!}{\includegraphics{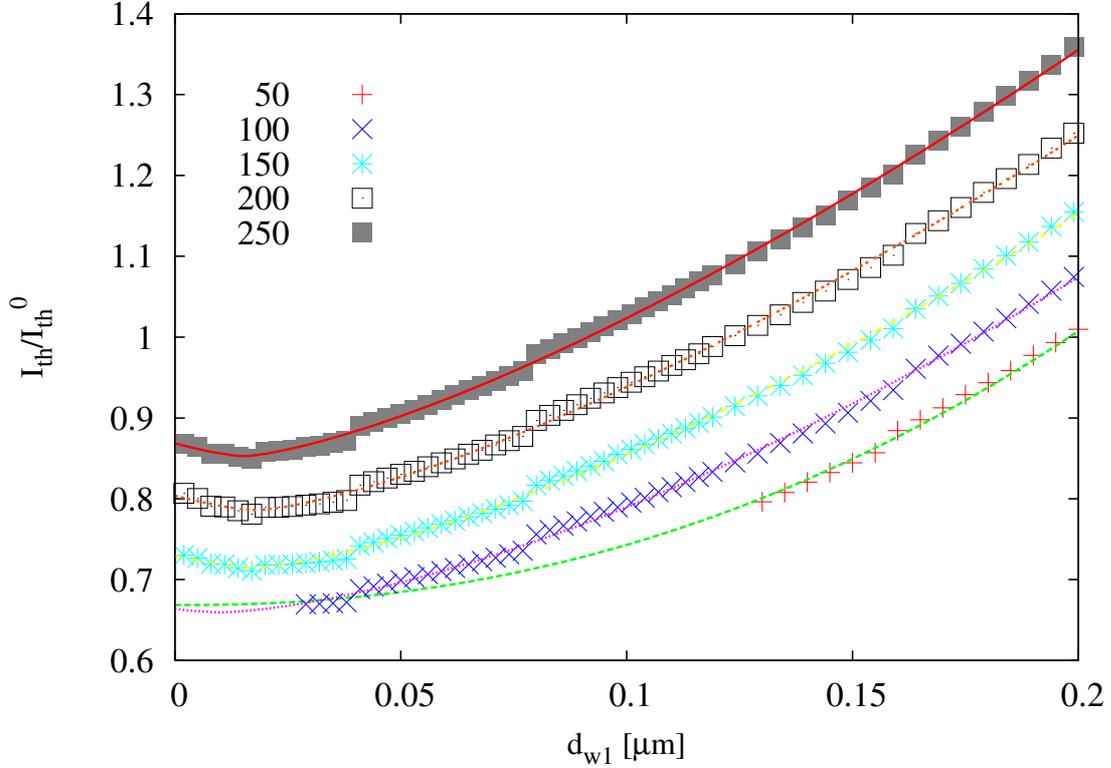}}
      \caption{Three Quantum Wells: threshold current, $I_{th}$, normalized by $I_{th}^0$, 
	as a function of the width of two waveguides in the center, for a few values 
	of other two waveguides width (in $nm$), as indicated in the Figure.
}
      \label{3active_00}
\end{center}
\end{figure}

\begin{figure}[h]
\begin{center}
      \resizebox{150mm}{!}{\includegraphics{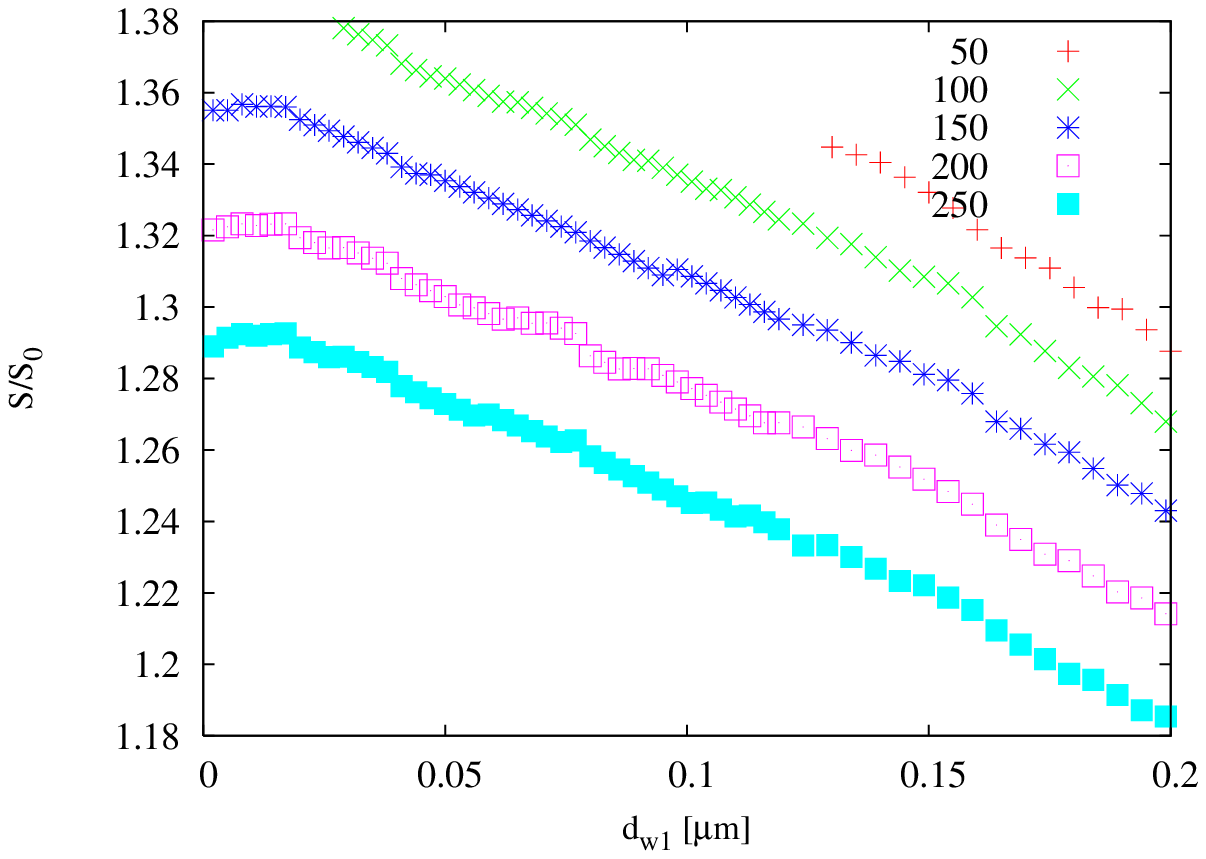}}
      \caption{Three Quantum Wells: Slope of optical power, $S=dL/dI$, normalized by $S_0$, 
	corresponding to the data in Figure \ref{3active_00}.
}
      \label{3active_01}
\end{center}
\end{figure}

Figures \ref{2active_00} and \ref{2active_01} show results for $I_{th}/I_{th}^0$ and $S/S_0$, respectively,
for the case of 2-QW structures, while Figures \ref{3active_00} and \ref{3active_01} show similar 
results for 3-QW structures. In these Figures, where no data points are present for certain values of
$d_{w1}$ - there is no lasing action observed there. The meaning of lines in Figures \ref{2active_00} 
\ref{3active_00} is to guide the eyes only. They are drawn however by using least-squares fitting
of the data points to simple power-law functions.

As we see, a significant reduction of threshold current is observed for both, 2-QW and 3-QW structures, 
if compared with the value for the reference laser with 1-QW. Also, a large increase of $dL/dI$ is found, 
more significant in case of 3-QW structure.

%%%%%%%%%%%%%%%%%%%%%%%%%%%%%%%%%%%%%%%%%%%%%%%%%%%%%%%%%%%%%%%%%%%%%%%%%%%%%%%%%%%%%%%%%%%
\section{Optical efficiency.}

\begin{figure}[h]
\begin{center}
      \resizebox{150mm}{!}{\includegraphics{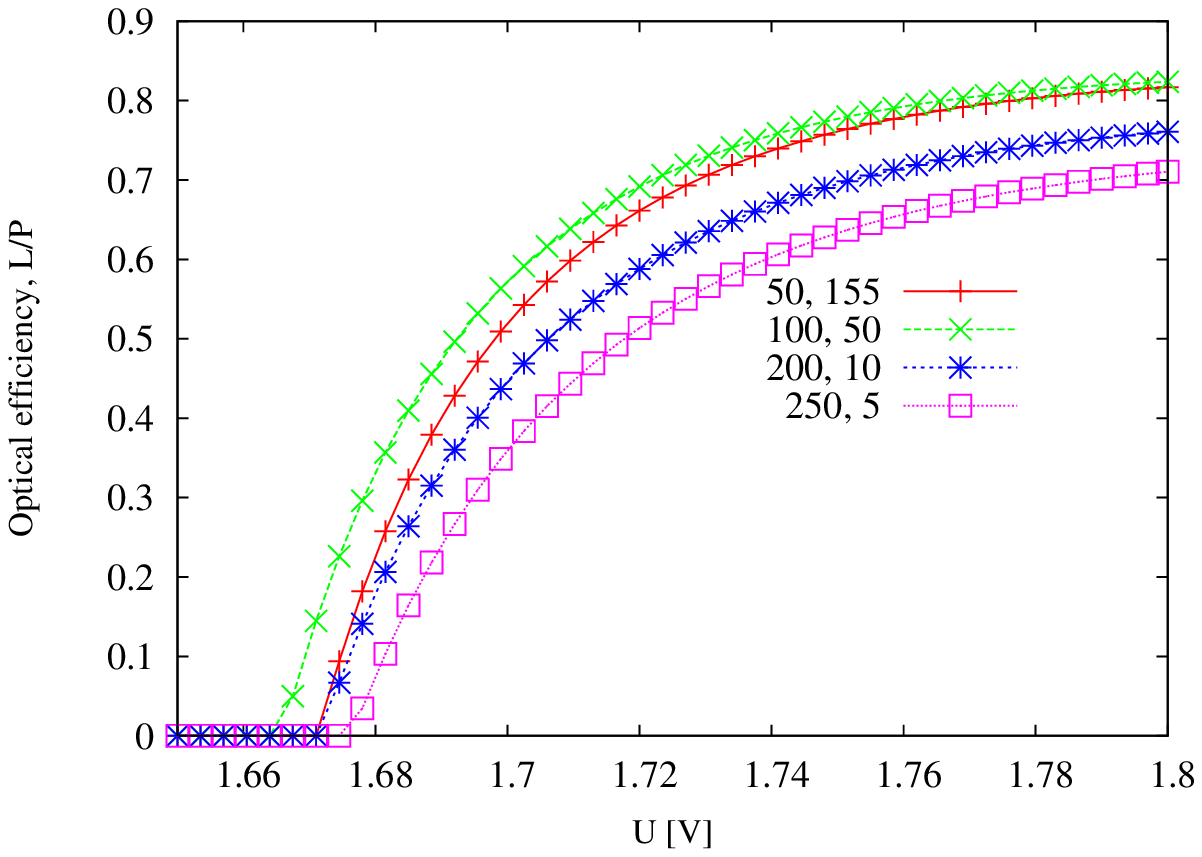}}
      \caption{Two Quantum Wells: Optical efficiency as a function of voltage applied
	for several combination of waveguides widths, as shown in the Figure:
	the first number is the width $d_{w0}$, the second one is $d_{w1}$ (in $nm$).
}
      \label{2active_02}
\end{center}
\end{figure}

\begin{figure}[h]
\begin{center}
      \resizebox{150mm}{!}{\includegraphics{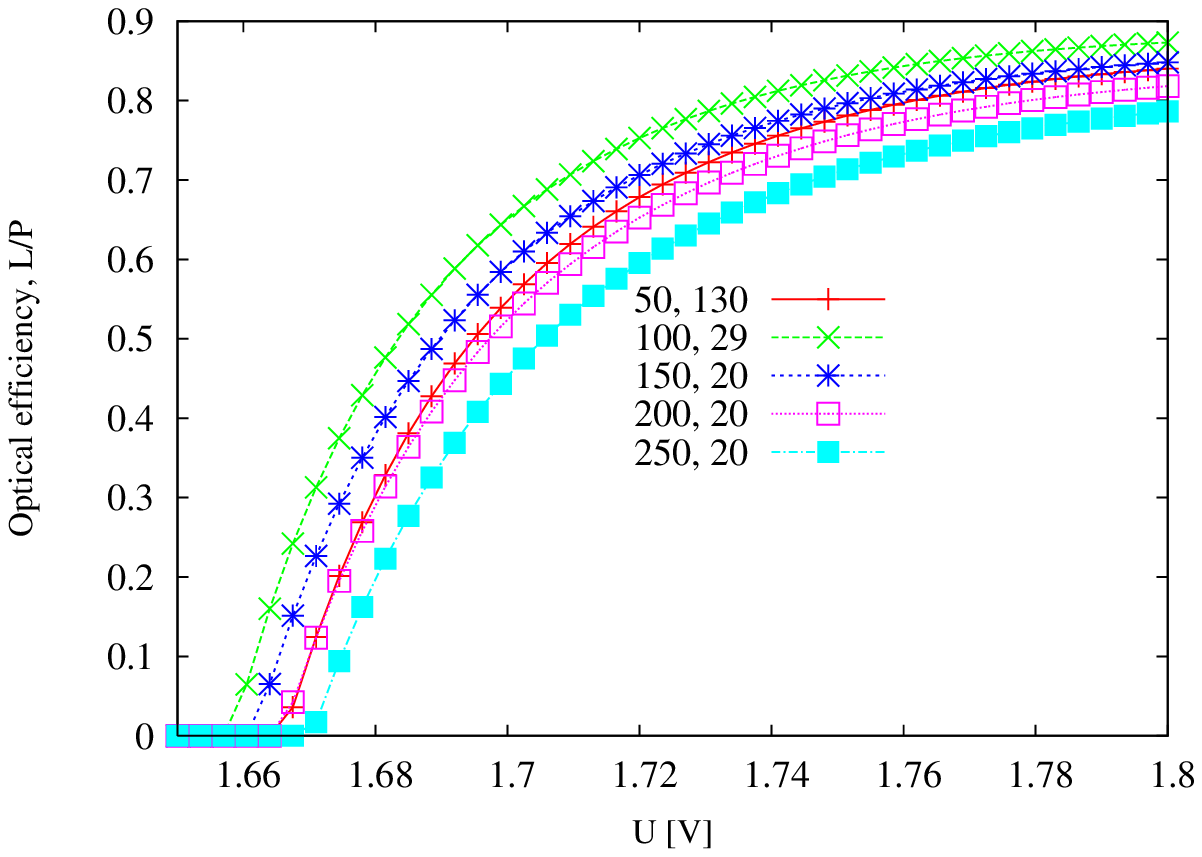}}
      \caption{Three Quantum Wells: Optical efficiency as a function of voltage applied
	for several combination of waveguides widths, as shown in the Figure:
	the first number is the width $d_{w0}$, the second one is $d_{w1}$ (in $nm$).
}
      \label{3active_02}
\end{center}
\end{figure}

Figures \ref{2active_02} and \ref{3active_02} illustrate how optical efficiency as a function of 
voltage applied depends on geometry of lasers, for several cases of waveguide parameters $d_{w0}$ and $d_{w1}$
that correspond, approximately, to the maximum values of optical efficiency. The largest optical efficiency 
achieved, of about 88\%, is for 3-QW structure when $d_{w0}$ is about 100 nm, and $d_{w1}$ of about 29 nm.

We proposed \cite{differential} that a modified exponential dependence describes very well
current-voltage characteristics just above the lasing offset $U>U_0$:

\begin{equation}\label{exponential_6_parameters}
\begin{array}{ll}
	I(U) = I_{th} \cdot exp(C\cdot (U-U_0) + D \cdot (U-U_0)^2),
\end{array}
\end{equation}

where $C$ and $D$ are certain fitting parameters. It is convenient to rewrite \ref{exponential_6_parameters} 
in dimensionless variables:

\begin{equation}\label{dimensionless}
\begin{array}{ll}
	i(u) = exp\left(\frac{1}{\alpha} \cdot (u-1) \cdot \left[ 1 + \beta \cdot U_0^2 \cdot (u-1) \right] \right),
\end{array}
\end{equation}

where we defined: $i(u)=I(U)/I_{th}$ and $u=U/U_0$, $\alpha= r \cdot I_{th} / U_0$, $\beta=U_0 \cdot D/C$, 
and we defined also $r=dU/dI=\frac{1}{I_{th} \cdot C}$, which is differential resistivity just above $U_0$.

Hence, the optical efficiency, $\eta=L/(U \cdot I)$, is given by:

\begin{equation}\label{dimensionless_eta}
\begin{array}{ll}
\eta(u) = \frac{S}{U_0} \cdot \frac {i(u)-1}{ u \cdot i(u)},
\end{array}
\end{equation}

where $i(u)$ is given by Eq. \ref{dimensionless}.

Like in case of 1-QW structure (\cite{differential}), the data 
shown in Figures \ref{2active_02} and \ref{3active_02} are described by \ref{dimensionless_eta} 
with a high accuracy, by using two fitting parameters only, $\alpha$ and $\beta$ (except of $I_{th}$ and $U_0$ that 
may be found in a straightforward way).

%%%%%%%%%%%%%%%%%%%%%%%%%%%%%%%%%%%%%%%%%%%%%%%%%%%%%%%%%%%%%%%%%%%%%%%%%%%%%%%%%%%%%%%%%%%
\section{Lasing offset voltage $U_0$.}

As shown by Eq. \ref{dimensionless_eta}, optical efficiency depends on lasing offset voltage
and it is desirable to have it's value as low as possible. In literature, there
is no a simple formula that would describe variations of $U_0$ with physical material
parameters and geometrical structure of a laser. One of reasons for that is that the 
$I-V$ characteristics depend strongly in a too complex way 
on density of states and number of captured carriers in QW. The number of QW levels and
their separation, as well their energy distance from conduction band or valance band offsets
are a function of QW geometry as well \cite{NGC}. Therefore, it is useful to have an insight
into what kind of dependencies might be obtained as a function of waveguide width. 
Performing real experiments of that kind is unrealistic: there is no way to control
laser geometrical dimensions during technological process with sufficient, required accuracy. 
Instead, a large spread of the data points would be obtained 
and no conclusions could be drawn\footnote{In \cite{NGC}, we point out however that there are
hypothetically ways that could bypass these experimental problems.}.

\begin{figure}[h]
\begin{center}
      \resizebox{150mm}{!}{\includegraphics{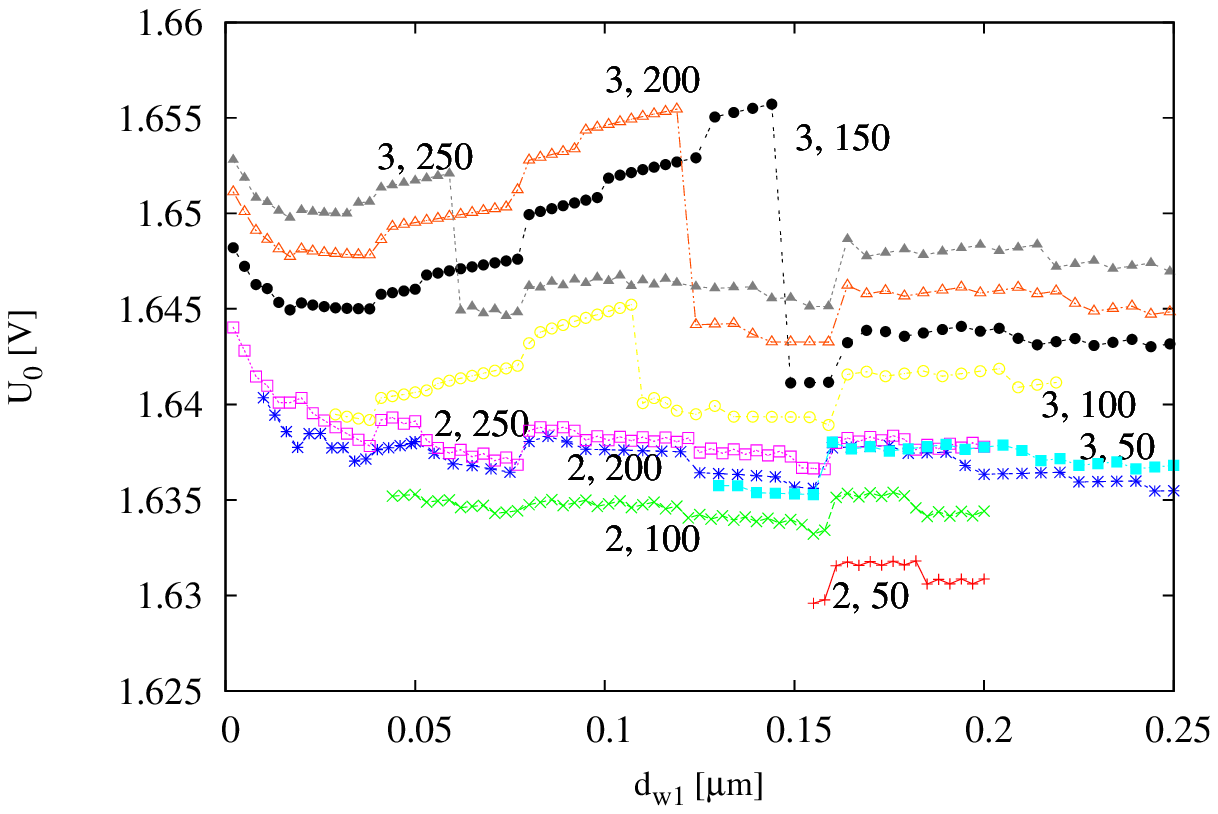}}
      \caption{Two and three Quantum Wells: lasing offset voltage $U_0$
		as a function of waveguide width $d_{w1}$. The first number in labels denotes 2- or 3-QW structure,
		the second one the width $d_{w0}$ in $nm$.
}
      \label{active_U0_00}
\end{center}
\end{figure}

We find in our modeling that a good accuracy of determining $U_0$ is gain versus voltage curve 
in a near range of voltage values below $U_0$. We use a linear extrapolation of the
data towards the maximum value, which is constant above $U_0$ in Sentaurus.
The results presented in Fig. \ref{active_U0_00} were obtained that way. This method
is the most accurate in our case and allows for easy, semi-automatic analysis 
of large collections of data sets.

As a result of a complex interplay between density of states in quantum wells 
when waveguide width $d_{w1}$ changes, clear discontinuities
are observed in $U_0(d_{w1})$, as shown on Fig. \ref{active_U0_00}.

One should however be cautioned here. Somewhat similar effects might be an undesired, artificial 
artifact of improperly conducting the modeling calculations: results depend for instance on
how the mesh changes when geometrical parameters of modeled structure change. In our case,
we kept the number of mesh points constant, independent of geometrical sizes of laser layers.

%%%%%%%%%%%%%%%%%%%%%%%%%%%%%%%%%%%%%%%%%%%%%%%%%%%%%%%%%%%%%%%%%%%%%%%%%%%%%%%%%%%%%%%%%%%%%%%
\section{Optical intensity profiles for 3-QW structure.}

When changing the structure and width of waveguides we gain some control over the power distribution
over the laser facets. That way we may possibly decrease the effects that limit the maximal output power,
that are caused by thermal damage to mirrors. 

One of the simplest ways to characterize optical intensity profiles is to use their half-width, i.e. 
the distance from the laser center where optical intensity decreases to half of its maximal value,
$\Delta w$.

\begin{figure}[h]
\begin{center}
      \resizebox{150mm}{!}{\includegraphics{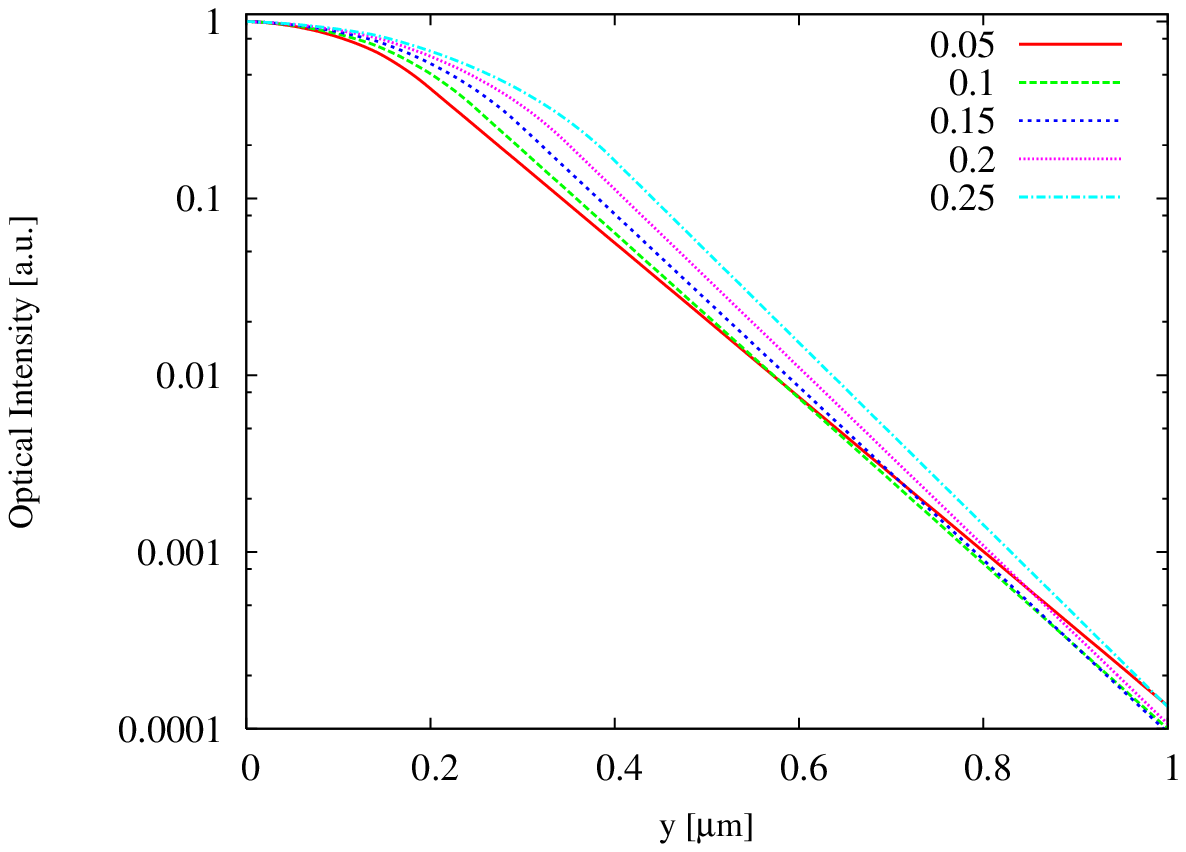}}
      \caption{Three Quantum Wells: Optical power intensity as a function 
	of distance from the laser center, $y$. Calculations were done 
	for the width $d_{w1}=0.130 \mu m$, and several values of $d_{w0}$ (in $nm$), as shown in the Figure.
	All curves are normalized to value of 1 at maxima.
}
      \label{3active_optical_01}
\end{center}
\end{figure}

The half-width in Figure \ref{3active_optical_01}, is given by linear 
dependence on $w_0$: $\Delta w = 0.409 \cdot w_0 + 0.161$ (when $w_1=0.130 \mu m$).

When similar analysis is done for the case of $w_1$ width of $0.029 \mu m$,
we obtain the following expression for the half-width of optical profile intensity:
$\Delta w = 0.394 \cdot w_0 + 0.111$.

For the case of $w_1$ width of $0.219 \mu m$, 
we obtain the following expression: $\Delta w = 0.444 \cdot w_0 + 0.2034$.

By additional analysis, we find that the following general relation describes
well the dependence of the half-width of optical intensity profile on $w_0$ and $w_1$:

\begin{equation}\label{DeltaW}
\Delta w = \left( 0.392 + 1.069 \cdot w_1^2\right) \cdot w_0 + 0.487 \cdot w_1 + 0.097
\end{equation}

where all quantities are in $\mu m$.
% Notice that in calculations, CosHyper broadaning is used, with parameter Broadaning =0.02.
% That value will probably affect the vlue of the last parameter in equation above.
% There will probably exist a contribution from QW width. In this case I would expect
% a contribution of value 0.018 (0.012 from one QW and 0.006 from half of QW in the middle.

%%%%%%%%%%%%%%%%%%%%%%%%%%%%%%%%%%%%%%%%%%%%%%%%%%%%%%%%%%%%%%%%%%%%%%%%%%%%%%%%%%%%%%%
\section{Summary and Conclusions.}

By using Synopsys's Sentaurus TCAD, and open source software we performed computer
modeling of optical and electrical characteristics of AlGaAs lasers with separate 
confinement heterostructures, when 2 and 3 quantum wells are present. We compared results 
with these for 1-QW laser calibrated to reproduce characteristics of lasers produced at Polyus
research institute in Moscow (\cite{Andrejev} and \cite{Andrejev_2}).

It was shown that a significant improvement of laser parameters is obtained in case of MQW lasers, 
when the width of waveguides is optimized.

The maximum optical efficiency achieved reaches 88 \% for a 3-QW structure. 

The width of optical intensity profile of MQW lasers increases, 
leading to lowering maximal light power density passing through laser facets, decreasing the 
risk of catastrophic damage of mirrors.

It has been shown, by examining lasing offset voltage, that laser characteristic parameters 
are expected to be discontinuous functions of waveguide's width.

We did not study possible more complex structures. In particular, as modeling
results show in case of 1-QW (\cite{Koziol_doping} and \cite{Waveguide}), 
introducing gradual doping profiles
in waveguides, and also gradual changes to Al concentrations there may lead to significant
improvement of laser parameters. Also, the assumed here width of quantum wells, is close
ony to the optimal one.

Moreover, it is tempting to create structures with variable width of active regions 
(but the same lasing frequency). These would have, for instance, different characteristics 
as a function of temperature.

\section{Acknowledgement}
This research was carried out under the Federal Program "Research and scientific-pedagogical 
cadres of Innovative Russia" (GC number P2514). The authors are indebted for valuable comments 
and discussions to A.~A.~Marmalyuk of Research Institute "Polyus" in Moscow.

\end{document}